\documentclass[referee]{aa} 
\usepackage{graphicx}
\begin{document}
   \title{Early Rebrightenings of X-ray Afterglows from Ring-Shaped GRB Jets}


   \author{M. Xu$^{1, 2}$
          \and
          Y. F. Huang$^{1, 2}$
          }

   \institute{$^1$ Department of Astronomy, Nanjing University,
              Nanjing 210093, China\\
              $^2$ Key Laboratory of Modern Astronomy and Astrophysics (Nanjing University),
              Ministry of Education, China \\
              \email{hyf@nju.edu.cn}
             }

   \date{Received 00 00, 0000; accepted 00 00, 0000}


  \abstract
    {}
   {Early rebrightenings at a post-burst time of $10^2$ --- $10^4$
s have been observed in the afterglows of some gamma-ray bursts
(GRBs). Unlike X-ray flares, these rebrightenings usually last
for a relatively long period. The continuous energy injection
mechanism usually can only produce a plateau
in the afterglow light curve, but not a rebrightening.
Also, a sudden energy injection can give birth to a rebrightening,
but the rebrightening is a bit too rapid.
}
  {Here we argued that the early rebrightenings can be produced by the ring-shaped jet model.
In this scenario, the GRB outflow is not a full cone, but a centrally hollowed ring.
Assuming that the line of sight is on the central symmetry axis of the hollow cone,
we calculate the overall dynamical evolution of the outflows and educe the multiband afterglow light curves.}
   {It is found that the early rebrightenings observed in the afterglows of a few GRBs,
such as GRBs 051016B, 060109, 070103 and 070208 etc, can be well explained in this framework. }
   {It is suggested that these long-lasted early rebrightenings in GRB
afterglows should be resulted from ring-shaped jets.}


\authorrunning{M. Xu \& Y. F. Huang}

\titlerunning{Afterglows from ring-shaped GRB jets}

   \maketitle
%

\section{Introduction}

   Gamma-ray bursts (GRBs) are powerful explosive events in the Universe.
The standard fireball model suggests that the prompt emission should result
from internal shocks and the afterglow should result from external shocks.
Since the launch of the $Swift$ satellite (Gehrels et al. 2004), enormous
improvements have been achieved in understanding the nature of GRBs
(for recent reviews, see: Zhang 2007, Gehrels et al. 2009).
Generally, some interesting components have been identified in the X-ray
afterglow light curves of GRBs, i.e., the steep decay phase,
the shallow decay phase, the normal decay phase, the post jet-break phase,
and X-ray flares.

Rebrightening behavior is also a very interesting feature among a few GRBs.
Unlike X-ray flares which are usually characterized by a rapid rise and fall
of the flux (with the mean ratio of width to peak time
$\langle\Delta t/t\rangle = 0.13 \pm 0.1$, Chincarini et al. 2007),
these rebrightenings are generally very gentle, with $\Delta t/t>1$.
Rebrightenings can occur either in the early afterglow
stage ($t \leq 10^4$ s) or in the late afterglow stage ($t \gg 10^4$ s).
Rebrightening in the late stage was first discovered in the X-ray and optical
afterglow of GRB 970508, at a post-burst time of about $t \sim 2$ day
(Piro et al. 1998). More examples of late rebrightenings
(typically occuring at $t \geq 1$ --- 2 day) can be found in
GRBs 030329 (Lipkin et al. 2004), 031203 (Ramirez-Ruiz et al. 2005),
050408 (de Ugarte Postigo et al. 2007), and 081028 (Margutti et al. 2009).
Interpretations for these late rebrightenings include off-axis jet model,
and microphysics variation mechanism (Kong et al. 2010).

In this study, we will mainly concentrate on the early rebrightenings
that typically occur at a post-burst time of $t \leq 10^4$ s.
Examples of such early rebrightenings can be found in GRBs 051016B,
060109, 070103, 070208 etc.
Most GRBs, like GRB 061007 (Kocevski \& Butler 2008), do not
exhibit early rebrightenings. Among more than 500 Swift GRBs
(Gehrels et al. 2009), we note that only less than 1\% events show early
X-ray rebrightenings. 
The energy-injection mechanism is a natural explanation
for rebrightenings, but it is unlikely to take effect here. For example,
the continuous energy-injection mechanism usually is more likely to
produce a plateau-like structure in the afterglow light curve,
but not an obvious rebrightening (Dai \& Lu 1998; Zhang \& M\'esz\'aros 2001;
Yu \& Huang 2007). Also, although a sudden energy-injection can give birth
to a rebrightening, the rebrightening is generally a bit too
rapid (Huang et al. 2006) as compared with the early rebrightenings
considered here.

We suggest that this kind of early rebrightenings may be produced by ring-shaped
GRB jets. A ring-shaped jet is not a full cone, but a centrally hollowed
ring (Granot 2005; Fargion \& Grossi 2006; Xu et al. 2008).
In this paper, we numerically investigate the afterglow features of such outflows
and use this mechanism to explain a few observed afterglows. The structure
of our paper is as follows. The geometry of ring-shaped jets is sketched in
Section 2. The dynamical evolution and afterglow of ring-shaped jets are
studied in Section 3, together with a detailed comparison with a few observations.
Finally, Section 4 is our discussion and conclusion.


\section{Ring-Shaped Jet}

A ring-shaped jet can be produced by the central engine during the
prompt GRB phase. The most natural mechanism may be via
precessing (Fargion \& Grossi 2006; Zou \& Dai 2006).
In almost all GRB progenitor models, the central engine is a
rapidly rotating compact objects, such as a black hole or a
neutron star. The outflows accounting for the GRB are usually
launched along the rotating axis or the magnetic
axis. If the nozzle of the central engine is
precessing while ejecting the outflow, then a ring-shaped jet
will be naturally produced. As an example, if GRBs are associated
with the kick of neutron stars (Huang et al. 2003), then the
production of ring-shaped jet will be quite likely.
Alternatively, the apparent ring-shaped topology of the
outflow can also arise from a structured jet as suggested by
van Putten \& Levinson (2003) and van Putten \& Gupta (2009).
This mechanism can explain the correlation between the observed
luminosity and variability in the light curves of long GRBs
reported by Reichert et al. (2001). 

A detailed sketch of the ring-shaped jets has been
presented in Xu et al. (2008). Here we will adopt the notations
there. The half opening angle of the inner edge
of the ring is denoted as $\theta_{\rm c}$, and the width of the ring
is $\Delta \theta$. Both the inner and outer edges may expand
laterally, so the inner edge will converge at the central axis
and the ring-shaped jet will finally become a conical jet after
a period of time. In our calculations, for simplicity, we assume
that the line of sight is on the symmetry axis of the centrally
hollowed cone. Therefore, this scenario is more or less
similar to that of a normal {\em off-axis} jet at early stages, i.e. the
line of sight is not on the emitting outflow. A rebrightening is
then expected to present in the early afterglow phase, thanks to the
deceleration and lateral expansion of the ring-shaped jet.

Since our line of sight is ``off-axis'' initially, a natural problem is
whether the ring-shaped jet could produce the intensive $\gamma$-ray
radiation in the prompt GRB phase, similar to the problem confronted
by a normal off-axis jet model. Below, we show that the ring-shaped
jet model has more advantages against the off-axis jet model as long as
this problem is concerned. The key point is that all the material on
a ring can contribute equally to the flux on the central axis.

Assuming that the bulk Lorentz factor of a highly relativistic point-like
jet is $\gamma \gg 1$, and that it is viewed from an angle $\Theta$
relative to its motion, then the observed flux will be amplified by a
factor of $\sim D^{3}$, due to the effect of relativistic beaming and Doppler
boosting. Here the Doppler factor is $D=[\gamma (1- \beta cos \Theta)]^{-1}$,
with $\beta=\sqrt{1-1/\gamma^2}$. For an extended outflow considered in
our framework, we should integrate over all the surface of the ring-shaped
jet to get the exact factor of amplification for the prompt GRB emission.

We first consider a normal conical jet with a full opening angle of
$\Delta \theta$ and a viewing angle of $\theta_{\rm c}+ \frac{1}{2} \Delta
\theta$ (similar to the viewing angle of our ring-shaped jet scenario).
The observed emission should be amplified by a factor of $f_{\rm j} =
\int_{\theta_{\rm c}}^{\theta_{\rm c} + \Delta \theta}D^{3}\phi(\Theta) d\Theta$
relative to that in the jet rest frame, where $\phi(\Theta)$ is the
toroidal angle given by
\begin{equation}
 \phi(\Theta)=2 \arccos [\frac{\cos(\Delta \theta /2) -
 \cos(\theta_{\rm c}+\Delta \theta /2) \cos\Theta}
{\sin(\theta_{\rm c}+\Delta \theta /2) \sin\Theta}].
\end{equation}
For simplicity, for a normal conical GRB jet with a full opening angle
of $\Delta \theta$ but with the viewing angle being $\sim 1/\gamma$,
we define the corresponding amplification factor as $f_{\rm o}$. Actually,
$f_{\rm o}$ is a measure of the amplification factor for an on-axis observer.

Now return to the ring-shaped jet considered in our framework. The
amplification factor for an observer on the central symmetry axis
is  $f_{\rm r}=\int_{\theta_{\rm c}}^ {\theta_{\rm c} + \Delta \theta}D^{3}
\phi(\Theta) d\Theta$, where $\phi(\Theta) = 2 \pi$.
To get a direct impression on this matter, let us simply take $\gamma=50,
\theta_{\rm c}=0.04$, $\Delta \theta=0.04$ as an example, and calculate the final
amplification factors. In this example, the viewing angle, i.e. the
angle between the jet axis and the line of sight,  equals $3/ \gamma$.
We can easily find that $f_{\rm o/r}\equiv f_{\rm o}/f_{\rm r}\approx 68.5$.
It means that the observed flux of our ring-shaped jet is only
tens of times less than that of an on-axis jet in the prompt GRB
phase. At the same time, we can also find that
$f_{\rm r/j}\equiv f_{\rm r}/f_{\rm j}\approx 12.4$. It tells us that the observed
flux of our ring-shaped jet is more than 12 times larger than that of
an off-axis conical jet. So, the ring-shaped jet model is superior
over the normal off-axis conical jet model when considering the
production of the intensive $\gamma$-ray radiation in the
prompt GRB phase.

\section{Numerical Results}
\label{sect:NCR}

The overall dynamical evolution of a ring-shaped jet has been studied
by Xu et al. (2008) in detail. Generally, if the line of sight is on
the central symmetry axis of the hollow cone, a rebrightening phase
should present in the early afterglow light curve. Here, for
completeness, we simply outline a few important ingredients of the
calculation. First, the dynamical equations should be appropriate
for both the ultra-relativistic and non-relativistic stages.
Recently, van Eerten et al. (2010) developed an accurate and also
correspondingly complex code for the dynamical evolution of
GRB afterglows. Here, we will use the simple and convenient
equations proposed by Huang et al. (1999). Secondly,
for a ring-shaped jet, both the inner and outer edges may expand
laterally. When the inner edge converges at the central symmetry
axis, the ring-shaped jet will become a normal conical jet,
and then we only need to consider the sideways expansion of the
outer edge. We suppose that the lateral expansion is at the comoving
sound speed ($c_{\rm s}$) approximately given by
\begin{equation}
\label{cs2} c_{\rm s}^2 = \hat{\gamma} (\hat{\gamma} - 1) (\gamma -
1) \frac{1}{1 + \hat{\gamma}(\gamma - 1)} c^2 ,
\end{equation}
where $\hat{\gamma} \approx (4 \gamma + 1)/(3 \gamma)$ is the
adiabatic index (Dai et al. 1999).
We have $\hat{\gamma}\sim 4/3$ and $c_{s}=c/\sqrt{3}$
in ultrarelativistic limit, and $\hat{\gamma}\sim 5/3$ and $c_{s}
=\sqrt{5/9}\beta c$ in nonrelativistic limit.
The effect of $c_{\rm s}$ on the light
curve of a ring-shaped jet has been discussed by Xu et al. (2008)
in detail. Generally speaking, lateral expansion tends to make the light
curve steeper and leads to an earlier jet break. 
Thirdly, to calculate the afterglow flux, we integrate the emission
over the equal arrival time surface defined by
\begin{equation}
\label{eqt3} \int \frac{1 - \beta \cos \Theta}{\beta c} dR \equiv t,
\end{equation}
within the jet boundaries (Huang et al. 2000).

In this section, we will calculate the overall dynamical evolution and
multiband afterglow of ring-shaped jets, and compare the numerical
results with the observations of some $Swift$ GRBs with early
rebrightenings, such as GRBs 051016B, 060109, 070103, and 070208.
In our calculations, we assume a standard cosmology with
$\Omega_{\rm M}=0.27$, $\Omega_{\Lambda}=0.73$, and with the Hubble
constant of $H_{0}=71~{\rm km}~{\rm s}^{-1} {\rm Mpc}^{-3}$.

\subsection{GRB 051016B}

GRB 051016B was triggered and located as a soft burst by $Swift$-BAT
at 18:28:09 UT on October 16, 2005 (Parsons et al. 2005).
The light curve of its prompt emission is shown in Figure 1a.
Its duration is $T_{90} = 4.0 \pm 0.1$s. The photon index
in $15-150$ keV is $2.38 \pm 0.23$, and the fluence is $(1.7 \pm 0.2)
\times 10^{-7} {\rm erg ~ cm}^{-2}$ (Barbier et al. 2005).
Lying at a redshift of $z=0.9364$ (Soderberg et al 2005),
the isotropic-equivalent $\gamma$-ray energy release is $E_{\rm \gamma,iso}
\sim 7.6 \times 10^{50}$ erg. The light curve of early X-ray afterglow
shows a rising phase beginning at about 200s, and the rebrightening lasts
for almost one thousand seconds.

Using the ring-shaped jet model, we have fit the observed X-ray and
optical afterglow light curves of GRB 051016B numerically. The result is
presented in Figure 2, and the parameter values are given in Table 1.
The observed X-ray afterglow data are relatively abundant.
We see that our model can reproduce the rebrightening and the overall
X-ray afterglow light curve satisfactorily. Note that in the very early
stage ($t \leq 200$ s), the afterglow may still be in the steep decay
phase and the flux should be dominant by the contribution from the prompt
tail emission. We thus have excluded these early data points from our
modeling. In Figure 2, the observed R-band optical afterglow light
curve is built up only by two data points, but they are in good agreement
with our prediction. From Table 1, we see that the ratio of $f_{\rm r/j}$ is
about 5.8. It means that in the prompt GRB phase, the $\gamma$-ray flux
of our model is about six times more than that of a corresponding
off-axis conical jet. So, the ring-shaped jet model is better than the
off-axis conical jet model when considering the production of the
intensive $\gamma$-ray radiation in the main burst phase.

   \begin{figure}[]
   \begin{minipage}[]{85mm}
   \centering
   \includegraphics[width=8.0cm, angle=0]{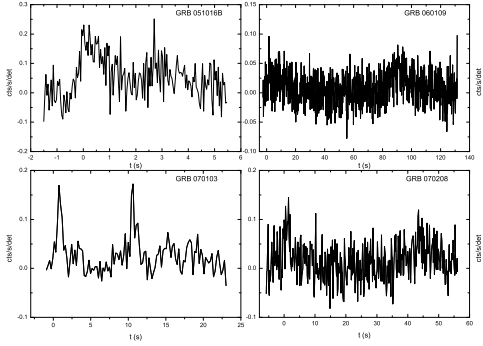}
   \caption{Prompt $\gamma$-ray light curves of GRBs 051016B,
   060109, 070103, and 070208 as observed by $Swift-BAT$ (15---350keV). The
   data are taken from the $Swift$ website\protect\footnotemark[1]
   (Butler \& Kocevski 2007; Butler et al. 2007). }
   \label{BAT}
   \end{minipage}
   \end{figure}

\footnotetext[1]{$http://astro.berkeley.edu/\sim nat/swift/$}

   \begin{figure}[]
   \begin{minipage}[]{85mm}
   \centering
   \includegraphics[width=8.0cm, angle=0]{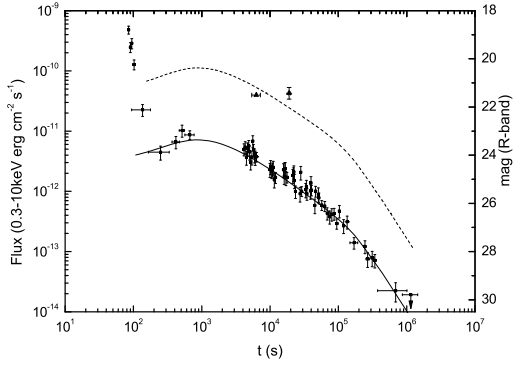}
   \caption{Our best fit to the X-ray (solid line) and R-band
optical (dashed line) afterglow light curves of GRB 051016B
by using the ring-shaped jet model. The square data points
are observed X-ray afterglow by $Swift$-XRT (see the $Swift$ website) and the triangle
data points are observed R-band afterglow (Chen et al. 2005;
Sharapov et al. 2005). }
   \end{minipage}
   \label{051016B}
   \end{figure}

\subsection{GRB 060109}

At 16:54:41 UT on January 9, 2006, $Swift$-BAT triggered and
located GRB 060109 (De Pasquale et al. 2006). The light curve
of its prompt emission is shown in Figure 1b. Its duration in
15 --- 350 keV is $T_{90} = 116 \pm 3$ s, and the fluence in
15 --- 150 keV band is $ (6.4 \pm 1.0) \times 10 ^{-7}
{\rm erg ~ cm}^{-2}$ (Palmer et al. 2006). No redshift is
measured for this event. If assuming a typical redshift of $z=1$,
then the isotropic-equivalent $\gamma$-ray energy release is
$E_{\rm \gamma,iso} \sim 3.74 \times 10^{51}$ erg. In the early X-ray
afterglow, the light curve shows a rebrightening that begins
in less than 1000 s and lasts for several thousand seconds.

Figure 3 illustrates our fit to the observed X-ray afterglow of
GRB 060109 by using the the ring-shaped jet model. The parameters
involved are given in Table 1. In our modeling, again we have omitted
the observed steep decaying phase. We see that the theoretical light
curve matches with the observational data very well.
The observed flux of the ring-shaped jet is about 10 times larger
than that of a corresponding off-axis conical jet ($f_{\rm r/j}\sim 10.4$,
see Table 1). The ratio of $f_{\rm o/r}$ is $\sim 48.5$. It rougly
equals to $E_{\rm K,iso} / E_{\gamma,iso}$, which is about 18 as can
be derived from Table 1. This proves that our modeling is self-consistent.

   \begin{figure}[]
   \begin{minipage}[]{85mm}
   \centering
   \includegraphics[width=8.0cm, angle=0]{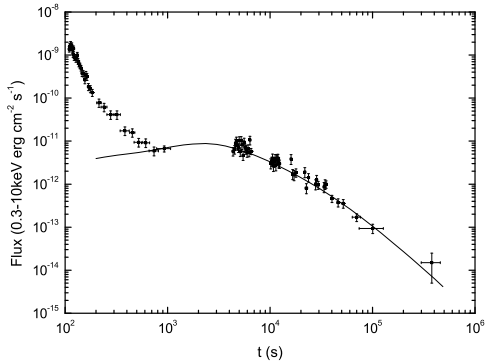}
   \caption{Our best fit to the X-ray afterglow of GRB 060109 by
using the ring-shaped jet model. The square data points are observed
by $Swift$-XRT (see the $Swift$ website). }
   \end{minipage}
   \label{060109}
   \end{figure}

\subsection{GRB 070103}

At 20:46:39.41 UT on 2007 January 3, $Swift$-BAT triggered and located
GRB 070103 (Sakamoto et al. 2007). The light curve
of its prompt emission is shown in Figure 1c. The duration in 15 --- 350 keV is
$T_{90} = 19 \pm 1$ s, and the fluence in 15 --- 150 keV is
$(3.4 \pm 0.5) \times 10 ^{-7} {\rm erg ~ cm}^{-2}$ (Barbier et
al. 2007). The isotropic-equivalent $\gamma$-ray energy release
is $E_{\rm \gamma,iso} \sim 1.8 \times 10^{51}$ erg, assuming
that the redshift is $z=1$. In the early X-ray afterglow, the light
curve shows a rebrightening that begins at about 200 s and lasts
for more than one thousand seconds.

In Figure 4, we illustrate our best fit to the X-ray afterglow
of GRB 070103 by using the ring-shaped jet model, with the involving
parameters given in Table 1. Generally, the observed rebrightening
and the overall light curve can be well explained. According to Table 1,
the value of $f_{\rm r/j}$ is about 5.3 . Note that the parameter of $f_{\rm o/r}$ is
about 52.2, and the ratio of $E_{\rm K,iso} / E_{\gamma,iso}$ is about 22.
Again these two values are roughly consistent with each other.

   \begin{figure}[]
   \begin{minipage}[]{85mm}
   \centering
   \includegraphics[width=8.0cm, angle=0]{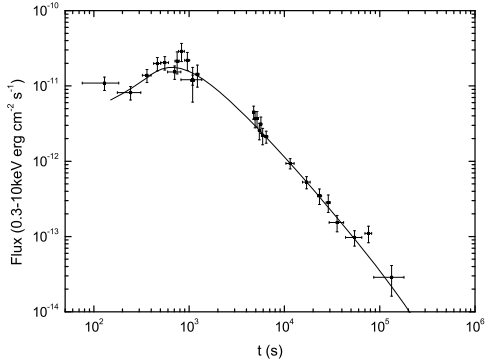}
   \caption{Our best fit to the X-ray afterglow of GRB 070103
by using the ring-shaped jet model. The square data points are
observed by $Swift$-XRT (see the $Swift$ website). }
   \end{minipage}
   \label{070103}
   \end{figure}

\subsection{GRB 070208}

GRB 070208 was triggered and located by $Swift$-BAT at 09:10:34 UT on
February 8, 2007 (Sato et al. 2007). The light curve
of its prompt emission is shown in Figure 1d. The duration in 15 --- 350 keV
is $T_{90} = 48 \pm 2$ s (Markwardt et al. 2007), and the fluence in
15 --- 150 keV is $(4.3 \pm 1.0) \times 10^{-7} {\rm erg ~ cm}^{-2}$.
Lying at a redshift of $z=1.165$ (Cucchiara et al. 2007), the
isotropic-equivalent $\gamma$-ray energy release is $E_{\rm \gamma,iso}
\sim 3.32 \times 10^{51}$ erg. The X-ray afterglow light curve shows a
rebrightening that begins at about 200 s and lasts for more than one
thousand seconds (Conciatore et al. 2007).

Using the ring-shaped jet model, we have tried to fit the multiband
afterglow of GRB 070208. The result is presented in Figure 5, with the
parameters given in Table 1. The observed early rebrightening in
the X-ray afterglow can be well reproduced. The optical afterglow
can also be simultaneously explained. Table 1 indicates that the
radiation intensity of the ring-shaped jet is about 8 times more
than that of a corresponding off-axis conical jet (i.e., $f_{\rm r/j}
\sim7.7$). Note that the value of $f_{\rm o/r}$ is about 61. It again
agrees well with the ratio of $E_{K,iso} / E_{\rm \gamma,iso}$,
which is about 24 in our modeling.

   \begin{figure}[]
   \begin{minipage}[]{85mm}
   \centering
   \includegraphics[width=8.0cm, angle=0]{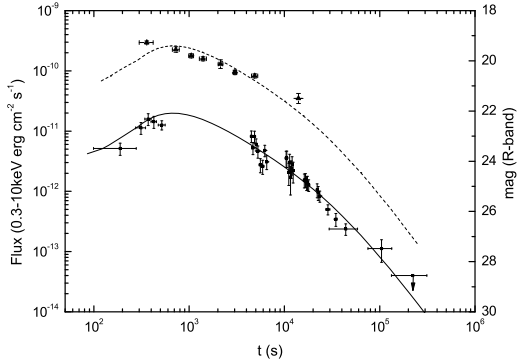}
   \caption{ Our best fit to the X-ray (solid line) and R-band
optical (dashed line) afterglow light curves of GRB 070208
by using the ring-shaped jet model. The square data points
are observed X-ray afterglow by $Swift$-XRT (see the $Swift$ website) and the triangle
data points are observed R-band afterglow (Cenko et al. 2009; Halpern \& Miraba 2007). }
   \end{minipage}
   \label{070208}
   \end{figure}

\section{Conclusion and Discussion}

It is possible that GRBs may be produced by ring-shaped jets.
Interestingly, a few recent hydrodynamical and magnetohydrodymical
simulations also give some supports to this idea
(Aloy \& Rezzolla 2006; Mizuno et al. 2008).
A notable feature of the afterglow of a ring-shaped jet is
that a rebrightening can be observed in the early afterglow stage,
assuming that the line of sight is within the central hollow cone.
In this study, we have clearly shown that the early rebrightenings
observed in a few GRBs, such as GRBs 051016B, 060109, 070103,
and 070208, can be well explained by the ring-shaped jet model.
In the case of GRBs 060109 and 070103, of which only X-ray afterglow
data are available, the fit result is very good. For GRBs 051016B
and 070208, of which both the X-ray and optical afterglow data are
available, the interpretation is also satisfactory. We propose that
these GRBs with long-lasting rebrightening in the early afterglow
should be produced by ring-shaped jets.

Comparing with a normal off-axis conical jet model, the advantage of our
ring-shaped jet model is obvious. For a normal conical jet, when the observer is
off-axis, the observed brightness will be notably reduced (as compared
to the on-axis case). It then has difficulty in explaining the
intensive $\gamma$-ray emission observed in the prompt
GRB phase. On the contrary, in the case of a ring-shaped jet, all the
material on the whole ring can contribute to the emission on the line
of sight. The observed intensity then can be significantly enhanced
as compared to that of an off-axis conical jet. In fact, for the 4
GRBs studied here, the amplification factor ($f_{\rm r/j}$)
is generally 5 --- 10, comparing with the corresponding off-axis
conical jet.

From Figure 1, we interestingly notice that all the four GRBs
show similar 2-pulse behavior in the prompt $\gamma$-ray
light curve.  This behavior is also observed in some other Swift GRBs
(Butler et al. 2007). It is quite unclear whether this 2-pulse behavior
is intrinsic to the mechanism that produces the ring-shaped jet or not.
Huang et al. (2003) have argued that if the jet is produced by rapid
precessing process (with the precession period much less than the GRB
duration), then the prompt $\gamma$-ray light curve could
be highly variable and very complicated. It then seems that these four
GRBs might not be due to normal fast precession. However, a slow precession
may still be possible. The structured jet mechanism by van Putten \&
Gupta (2009) is of course another choise. Other possibilities include
the evolution of an extreme Kerr Black Hole surrounded by a precessing disk
(Lei et al. 2007), or explanation of the second pulse as the tail emission
from the first pulse. 

From Table 1, we see that the initial angular radius of the inner edge of the
ring is generally very small ($\theta_{\rm c} \sim $ 0.01 --- 0.02).
The width of the ring is also small ($\Delta \theta \sim$ 0.02 --- 0.08).
Taking typical parameter values of $\theta_{c}=0.015$ and
$\Delta\theta=0.04$, we can give a rough estimate for the observed
event rate of early rebrightenings. Among all Swift GRBs,
the fraction of well-monitored afterglows with potential jet-breaks is
around $60 \%$ (Panaitescu 2007). For most of these jet candidates,
the outflows might be normal conical jets, but it is quite likely that a
small portion (here we take the ratio as 15\%, see Xu et al. 2008) of
the jets are ring-shaped ones. For a ring-shaped jet with
$\theta_{c}=0.015$ and $\Delta\theta=0.04$,
the possibility that our line of sight is within the central hollow cone
(as compared with the possibility that the line of sight is just on the
ring) is about 0.08 (calculated from
$(1-cos \theta_{c})/[(1-cos(\theta_{c}+\Delta\theta))-(1-cos\theta_{c})]$).
Final, we estimate that the fraction of Swift events predicted by our model
to display early rebrightening is $0.6\times0.15\times0.08 \sim 0.7\%$.
This number is consistent with the observed fraction of Swift events
that show early X-ray rebrightening ($<1\%$).

The angular information may provide useful clues on the central engine of GRBs.
In many progenitor models of GRBs, the central engine is a rapidly
rotating compact star and the outflow is ejected along the magnetic
axis. Our study then indicates that the inclination angle of the magnetic
axis with respect to the rotating axis should be small ($\sim$ 0.02 --- 0.06).
In a few other progenitor models, the central engine is also a rapidly
rotating compact star, but the outflow is ejected just along the rotating
axis, then our results suggest that the processing (angular) radius of the
rotating axis should be $\sim$ 0.02 --- 0.06.

In our framework, since the rebrightening is mainly a geometric effect,
it should generally be achromatic, i.e. the rebrightening should appear
simultaneously in all the X-ray and optical bands.  For the four GRBs
studied here, the X-ray afterglow data are relatively prolific. But the
optical afterglow data are generally so lacking that almost no rebrightening
can be discerned in the optical light curves. Anyway, we have shown that
our model is consistent with both the X-ray and optical observations.
In the future, more examples with abundant multiband afterglow data will
be available, and tighter constraints on the existence of ring-shaped jet
could be derived.

In our model, we assume the ring-shaped jet is uniform. But note
that the actual structure of the GRB outflow may be very complicate.
For example, it may be two-component jet as predicted by some engine
models (e.g. van Putten \& Levinson 2003). If these ingredients are
included, then the afterglow behavior of ring-shaped jets will be
correspondingly much more complicated.

\begin{acknowledgements}
We thank Fayin Wang and Yang Guo for helpful discussion.
We also would like to thank the anonymous referee for useful comments
and suggestions that led to an overall improvement of this manuscript. 
This work was supported by the National Natural Science Foundation of
China (Grant No. 10625313) and the National Basic Research Program of
China (973 Program, Grant No. 2009CB824800).
\end{acknowledgements}

\begin{table}
\begin{minipage}[]{120mm}
\centering
\caption[]{Parameters of the 4 GRBs}

\small
 \begin{tabular}{ccccccccccc}
  \hline\noalign{\smallskip}
GRB Name & GRB 051016B & GRB 060109 &  GRB 070103  &
GRB 070208 \\
  \hline\noalign{\smallskip}
z   & $0.9364$ &  $1$ &  $ 1 $ & $1.165$ \\
$E_{\rm \gamma,iso}$ (erg) & $7.6\times10^{50}$ &  $3.37\times10^{51 }$ &  $ 1.79\times10^{51 }$ & $3.32\times10^{51 }$ \\
$E_{\rm K,iso}$ (erg) & $1.3\times10^{52}$ &  $6.0\times10^{52 }$ &  $ 4.0\times10^{52 }$ & $8.0\times10^{52 }$ \\
$\theta_{\rm c}$ (rad) & $0.01$ &  $0.018$ &  $ 0.01$ & $0.009$ \\
$\Delta \theta$ (rad) & $0.07$ &  $0.024$ &  $ 0.08$ & $0.022$ \\
$\gamma$ & $180$ &  $90$ &  $ 120$ & $150$ \\
n $(cm^{-3})$ & $0.05$ &  $0.08$ &  $ 0.1$ & $0.1$ \\
$\epsilon_{\rm e}$ & $0.1$ &  $0.1$ &  $ 0.1$ & $0.1$ \\
$\epsilon_{\rm B}$ & $5.0\times10^{-4}$ &  $1.0\times10^{-4}$ &  $1.0\times10^{-4}$ & $1.0\times10^{-4}$ \\
$p$ & $2.05$ &  $2.2$ &  $ 2.75$ & $2.02$ \\
$f_{\rm r/j}$ & $5.8$ &  $10.4$ &  $5.3$ & $7.7$ \\
$f_{\rm o/r}$ & $228.7$ &  $48.5$ &  $52.2$ & $61.0$ \\
  \noalign{\smallskip}\hline
\end{tabular}

\begin{list}{}{}
\item[] \textbf{Notes:} ~~ z: {\sl redshift};~~ $E_{\rm \gamma,iso}$: {\sl the
isotropic-equivalent $\gamma$-ray energy release observed by $Swift$};
~~ $E_{\rm K,iso}$: {\sl the isotropic-equivalent kinetic energy
of the outflow used in our model};~~ $\theta_{\rm c}$: {\sl the half opening
angle of the inner edge of the ring};~~ $\Delta \theta$: {\sl the
width of the ring};~~ $\gamma$: {\sl the bulk Lorentz factor};~~
$n$: {\sl number density of the circum-burst medium};~~ $\epsilon_{\rm e}$: {\sl the
electron energy fraction};~~ $\epsilon_{\rm B}$: {\sl the magnetic energy
fraction};~~ $p$: {\sl the power-law index of the energy distribution
function of electrons};~~ $f_{\rm r/j}$: {\sl $f_{\rm r}/f_{\rm j}$ (see Section 2
for details)};~~ $f_{\rm o/j}$: {\sl  $f_{\rm o}/f_{\rm r}$ (see Section 2 for details).}
\end{list}
\end{minipage}

\end{table}

\end{document}